\documentclass{article}
\usepackage{spconf,amsmath,graphicx, amsfonts, bm}
\usepackage{multirow}
\usepackage{booktabs}
\usepackage[font=small,labelfont=bf]{caption}

\usepackage{fancyhdr}

\fancypagestyle{plain}{%
    \fancyhf{}%
    \fancyfoot[L]{\footnotesize \center \emph{Preprint accepted for publication in the Proceedings of IEEE Automatic Speech Recognition and Understanding Workshop (ASRU 2019). Singapore. 2019. \\ -\thepage-}}%
}
\newcommand{\ctx}{\mathbf{c}}
\newcommand{\E}{\mathbb{E}}
\newcommand{\I}{\mathcal{I}}
\newcommand{\Loss}{\mathcal{L}}
\newcommand{\N}{\mathcal{N}}
\newcommand{\m}{\mathbf{\mu}}
\newcommand{\R}{\mathcal{R}}
\newcommand{\s}{\bm{\sigma}}
\newcommand{\X}{\mathbf{X}}
\newcommand{\z}{\mathbf{z}}
\newcommand{\y}{\mathbf{y}}

\title{singing voice conversion with disentangled representations of \\singer and vocal technique using variational autoencoders}
\name{Yin-Jyun Luo$^{1, 3}$ \qquad
      Chin-Cheng Hsu$^{2}$ \qquad
      Kat Agres$^{3, 4}$ \qquad
      Dorien Herremans$^{1,3}$\thanks{This work is supported by a SINGA provided by the A*STAR, under reference number SING-2018-01-1270.}}
\address{
    $^1$ Singapore University of Technology and Design\\
    $^2$  University of Southern California, Los Angeles, United States\\
    $^3$ Institute of High Performance Computing, A*STAR, Singapore\\
    $^4$ Yong Siew Toh Conservatory of Music, National University of Singapore}
\begin{document}
\pagestyle{fancy}
%
\maketitle
\begin{abstract}
We propose a flexible framework that deals with both singer conversion and singers vocal technique conversion. The proposed model is trained on non-parallel corpora, accommodates many-to-many conversion, and leverages recent advances of variational autoencoders. It employs separate encoders to learn disentangled latent representations of singer identity and vocal technique separately, with a joint decoder for reconstruction. Conversion is carried out by simple vector arithmetic in the learned latent spaces.
Both a quantitative analysis as well as a visualization of the converted spectrograms show that our model is able to disentangle singer identity and vocal technique and successfully perform conversion of these attributes.
To the best of our knowledge, this is the first work to jointly tackle conversion of singer identity and vocal technique based on a deep learning approach.

\end{abstract}

\section{Introduction}
\label{sec:intro}

Singing voice conversion (SVC) comprehensively refers to
tasks that convert an attribute of singing.
Converting from one singer's voice to that of another  without affecting linguistic content 
has been the focus in SVC \cite{villavicencio2010applying, doi2012singingvoicconversion, kobayashi2014statistical, nachmani2019unsupervised, chen2019singing}.
Converting between different vocal techniques,
however, is a worthwhile
line of research that has lacked attention.
Such an approach would allow one to convert a singing voice into a different timbre or pitch that was originally infeasible due to
physical constraints or lack of singing skills,
thereby facilitating applications in entertainment and pedagogy.

Vocal techniques, such as `breathy' and `vibrato',
enrich the sound and are an integral part of singing. 
Singers 
perform different techniques at different points in time so as to create emotional ebbs and flows.
Modeling vocal techniques through data-driven models is challenging 
due to lack of labeled and balanced data, together with intrinsic ambiguity, just to name a few.

\begin{figure}[t]
\centering
\includegraphics[width=7.3cm]{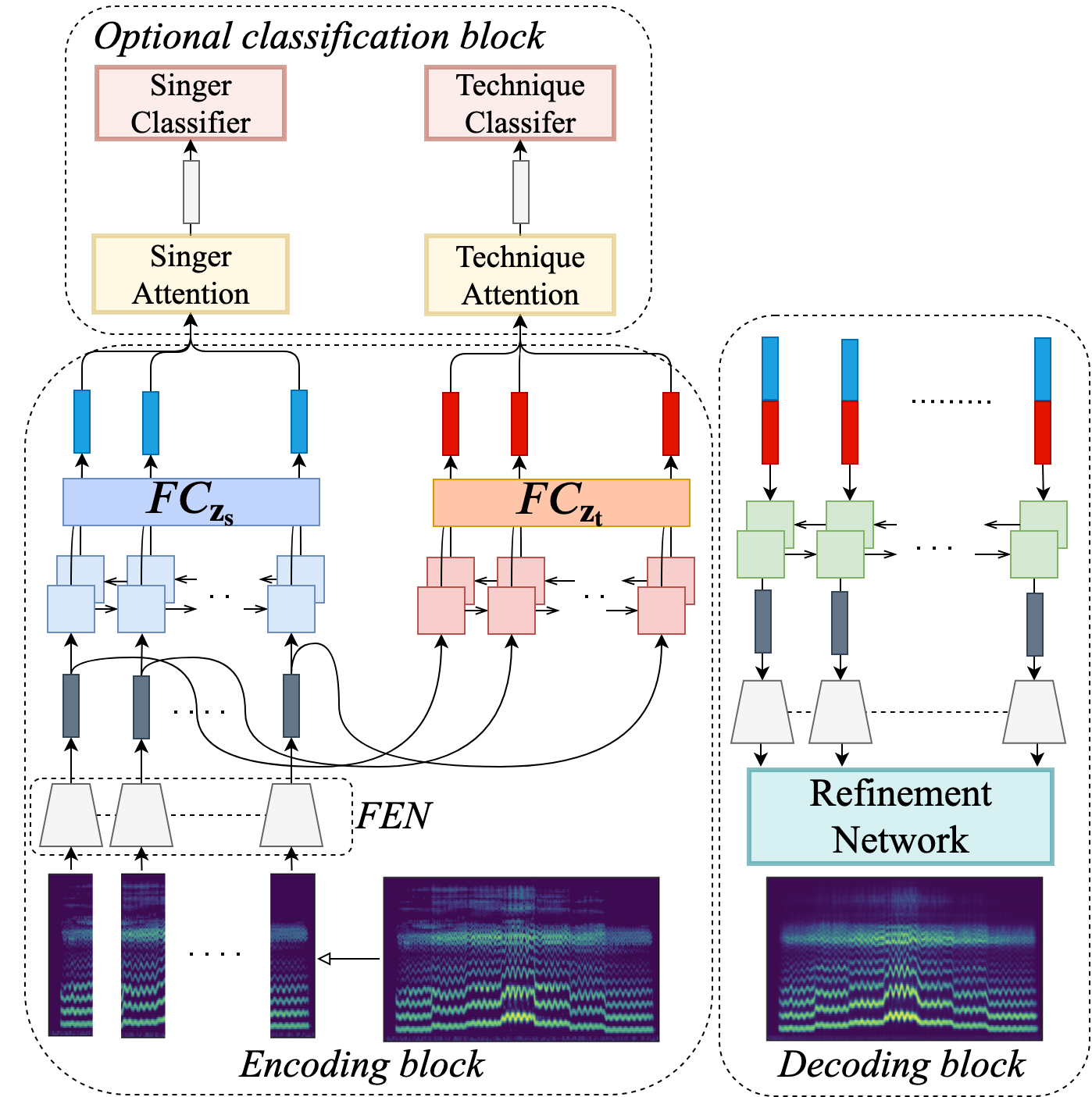}
\caption{The proposed framework, fully detailed in Section~\ref{sec:arch}. The blue, red and green blocks correspond to the singer encoder, vocal technique encoder, and the joint decoder, respectively.}
\label{fig:arch}
\end{figure}

We propose a framework that deals with the conversion of both singer identity and vocal technique. We augment the model based on the Deep Bi-directional Long Short-Term Memory (DBLSTM) from \cite{sun2015dblstm} with latent variables, such that it learns disentangled representations for both singer and vocal technique through Gaussian mixture variational autoencoders (GMVAEs) \cite{dilokthanakul2016deep, jiang2016variational}. 
Unlike typical SVC models that condition generation of singing voice on an utterance-level singer label \cite{sisman2019singan, chen2019singing, nachmani2019unsupervised}, our model is conditioned on time-dependent  singer/technique variables on a shorter temporal scale, accommodating cases in which the vocal technique varies across time. The proposed model can be trained on non-parallel (i.e., unpaired) corpora, and allows for many-to-many conversion of singer identities and vocal techniques.

We describe our modified implementation of the GMVAE model
\cite{hsu2019hierarchical, luo2019disentangle}
along with our singer/vocal technique conversion strategy in Section~\ref{sec:method}.
Next, we elaborate on our experimental setup using VocalSet
\cite{DBLP:conf/ismir/WilkinsSWP18},
a dataset featuring signing techniques, 
in Section~\ref{sec:settings}. 
Finally, we report and discuss the experimental results in Section~\ref{sec:results}.

\section{Method}
\label{sec:method}

\subsection{Variational Autoencoders}
\label{sec:vae}
Our proposed singing voice generation process ($\z \rightarrow \X$) based on VAE includes the generation of a chunk of spectrogram $\X \in \R^{T \times F}$ which is generated from 
a latent variable $\z \in \R^D$.
This simple dependency structure allows us to apply variational inference, which optimizes the evidence lower bound (ELBO) of $p(\X)$:
\begin{equation}\label{vae_elbo}
    \Loss(p, q; \X) 
    = \E_{ q( \z | \X)}[ \log p(\X | \z) ] 
    -D_{KL} ( q( \z | \X) || p(\z) ),
\end{equation}
where we assume
$p(\z) = \N(0, I)$,
$p(\X | \z) = \N(\m_x, \I)$ and
$q(\z | \X) = \N(\m_z, diag(\s_z))$,
in which $\m_z$ and $\s_z$ are inferred from $\X$ by an encoder, 
and $\m_x$ is predicted by a decoder.
Reconstructing $\X$ from $\z$ which is inferred from $\X$ itself using variational methods thus concludes a VAE.

\subsection{Increasing Expressivity using a GMM Prior}
\label{sec:gmvae}
The above made assumption $p(\z) = \N(0, I)$ reflects the preference for a simple distribution of data, which in turn sacrifices model expressivity. Replacing $p(\z)$ with a Gaussian mixture model (GMM), known as GMVAEs, has been proven effective in increasing the expressivity and controllability \cite{dilokthanakul2016deep,  
jiang2016variational,  hsu2019hierarchical, luo2019disentangle}.
It has an additional layer
of dependency: $\y \rightarrow \z$
that enables us to utilize categorical attributes $\y$ 
that may be available in data.
The ELBO of a GMVAE then becomes:
\begin{equation}\label{gmvae_elbo}
\begin{split}
    \Loss(p, q; \X, \y) 
    = \E_{q( \z | \X)}[ \log p(\X | \z) ] 
    -D_{KL}(q( \z | \X) || p(\z | \y)),
\end{split}
\end{equation}
where the prior $p(\z)$ is now multi-modal (GMM), more likely to model data with higher diversity. 
In addition, introducing $\y$ endows the model with direct controllability over attributes and flexibility for generation, as will be elaborated next.

\subsection{Controlling Singer Identities and Vocal Techniques}
\label{sec:ysyt}
Using the proposed GMVAE, the generation process for singing voice is as follows:
given a singer $\y_s$ and a vocal technique $\y_t$ (collectively referred to as \textit{attributes}),
the model first infers latent representations ($\z_s$ and $\z_t$) of each of the attributes, and then combines these two to generate a spectrogram of the singing voice.
Mathematically, the joint probability, given the attributes can be factorized as follows:
\begin{equation}
\begin{split}
    p(\X, \z_s, \z_t | \y_s, \y_t) = p(\X | \z_s, \z_t) p(\z_s | \y_s) p(\z_t | \y_t),
\end{split}
\end{equation}
where both the conditional distributions $p(\z_s| \y_s)$ 
and $p(\z_t | \y_t)$ are assumed to be
Gaussian with learnable means and diagonal covariances.
This GMVAE model now takes into consideration singer and vocal technique and thus can directly control them during the conversion phase.

\subsection{Learning an Attribute-discriminative Space}
\label{sec:att-classifier}
We incorporate two classifiers, 
one for vocal techniques and the other for singers,
to encourage the learned spaces $\z_t$ and $\z_s$ to be discriminative.
Each classifier learns to predict $\y_{*}$ from the sequence 
of $\z_{*} = \{\z_{*1}, \z_{*2}, ..., \z_{*N} \}$,
where $N$ is the number of chunks of a recording
and $*$ denotes singers or vocal techniques.
The classifier receives a sequence level representation that can be summarized by a simplified attention mechanism \cite{sonderby2015convolutional, collin2015feedforward}:
\begin{equation}\label{eq:attention}
    \alpha_{*n} = \frac{e^{f(\z_{*n})}}{\sum_{m=1}^{N} e^{f(\z_{*m})}},
    \ctx_{*} = \sum_{n=1}^{N}{\alpha_{*n} \z_{*n}},
\end{equation}
where $f(\cdot)$ is a learnable function 
and $\ctx_{*}$ denotes the summarization (and thus the representation) of the input sequence $\X$. $\alpha_{*n}$ is uniformly distributed without the attention module. The auxiliary objective of maximizing $p(\y_* | \X)$ will thus be added to the overall objective.

\subsection{Weighting KLDs}
\label{sec:kld-weight}
A singer might be unable to maintain the same level of expressiveness of a technique throughout a recording.
Similarly, voice timbre and pitch also vary across a recording even though a singer is asked to perform the same vocal technique. Based on these observations, we may benefit from weighting the KLD terms of \eqref{gmvae_elbo} with $\alpha_{*n}$ obtained from Section~\ref{sec:att-classifier}. Consummating our training objective, we have
\begin{equation}
\label{eq:objective}
\begin{split}
    \Loss(p, q; \X, \y_s, \y_t) = 
    &\E_{q(\z_s | \X) q(\z_t | \X)} \big[ \log p(\X | \z_s, \z_t)  \big] \\
    - \sum_{n=1}^{N} &\alpha_{tn} D_{KL}( q(\z_{tn} | \X) ||  p(\z_t | \y_t) ) \\
    - \sum_{n=1}^{N} & \alpha_{ts} D_{KL}( q(\z_{sn} | \X) ||  p(\z_s | \y_s) ) \\
    + \beta \log &p(\y_s | \X) + \gamma \log p(\y_t | \X),
\end{split}
\end{equation}
where $\beta$ and $\gamma$ are weights for the discriminative objectives.

\subsection{Conversion Strategies} 
\label{sec:conversion_strat}
We accommodate the time-varying singing attributes in expressive singing voices by learning latent variables at a shorter temporal scale.
Consequently, we adopt the model to infer the attributes,
rather than assigning attributes directly during the conversion phase.\footnote{Empirically, we found training a model conditioned on an utterance-level technique label did not work well, and we resorted to the proposed method.}
Generally, conversion is done by adding a conversion vector $d\m_{*n}$ to $\z_{*n}$ at the chunk level. We define $d\m_{*n} = \m_{*}^{target} - \m_{*n}^{source}$, where $\m_{*}$ represents the mean of a Gaussian mixture component. $\m_{*n}^{source}$ can be determined by either Gaussian likelihood $p(\y_{*n} | \z_{*n})$ or the auxiliary sequence-level classifier $p(\y_{*} | \X)$, referred to as \textit{C-chunk} and \textit{C-sequence}, respectively. Note that the latter computes a common $\m_{*n}^{source}$ (and hence $d\m_{*n}$) over all $N$ chunks. We report the result from both methods below.

\section{Experimental Settings}
\label{sec:settings}

\subsection{Dataset}
\label{sec:dataset}
We use the VocalSet~\cite{DBLP:conf/ismir/WilkinsSWP18} to evaluate our framework.
A subset of audio files we selected has 20 singers, 6 vocal techniques that were most distinguishable, (straight, breathy, vibrato, belt, lip trill and vocal fry), and 5 vowels. Each recording was sung as either a scale or arpeggios.
The subset was divided into a training set of 1,065 recordings, and a testing set of 118 (17 out of the $20 \times 6 \times 2 \times 5=1200$ combinations are missing). The length of recordings ranges from 3.5 to 23 seconds. This subset approximates a balanced number of instances over classes for singers, techniques, and vowels.

We re-sampled the recordings to 22,050 Hz, 
normalized the waveform w.r.t. the largest magnitude,
computed the log magnitude Mel-spectrogram (MEL) with $F=96$ filter banks, and then rescaled it to [-1, 1].
We further segmented the MEL into chunks of $T=43$ frames ($\X \in \R^{43 \times 96}$).
A frame shift of 256 was used for computing the MELs, so that
43 frames amount to 0.5 seconds.

\subsection{Architecture}\label{sec:arch}
Our SVC framework encompasses seven components: 
a feature extraction network (FEN), 
two encoders,
a decoder,
a post-processor,
and two sequence-level classifiers.
The overall architecture is shown Fig. \ref{fig:arch}

The FEN is composed of a two-layer one-dimensional convolutional neural network (CNN), 
each with 512 filters ($3\times1$), 
followed by two fully-connected (FC) layers with 512 and 256 units respectively. 
Batch normalization followed by \texttt{ReLu} is used for every layer. 
The FEN produces a 256-dimension bottleneck feature for a given input MEL chunk, 
and is shared and consumed by both encoders that follow.

Both of the encoders are parameterized as two-layer Recurrent Neural Networks (RNNs): a BLSTM 
with 256 hidden units, followed by two FC layers
shared across time which predict $\m_{*}$ and $\log \s_{*}$.
$\z_{*}$ is then sampled using the reparameterization trick from 
\cite{DBLP:journals/corr/KingmaW13}.
The joint decoder has a similar architecture as the encoders
but in reverse order. 
At each time step of the output sequence, 
a CNN with an architecture that is symmetric to the FEN is employed to reconstruct the MEL chunk. 
Finally, the refinement network, 
a three-layer one-dimensional CNN with 512 filters ($3\times1$), is used to refine the reconstructed MELs.

\subsection{Hyperparameters}
The mean vectors of $p(\z_{*}|\y_{*})$ were all randomly initialized, whereas the variance vectors were kept fixed with value $e^{-2}$. The number of mixtures for singers is set to 20, and set to 6 for vocal techniques. Both are equal to the number of classes they correspond to. We set the batch size to 128, and initialized the model parameters with Xavier initialization \cite{DBLP:journals/jmlr/GlorotB10}. The Adam optimizer \cite{DBLP:journals/corr/KingmaB14} was used
with a learning rate of $10^{-4}$.

\subsection{Evaluation Metrics}

We evaluate our model by how well a classifier correctly
recognizes the attributes in the converted MELs. 
The main idea is that the converted MELs should be accurately 
classified as the target class, 
and attributes that are not intended 
to be converted should be predicted the same. 
The classification results thereby reveal the effects 
on the output MELs caused by conversion.
Each recording in the test set is first 
converted to all possible target attributes 
and then evaluated by three classifiers 
that recognize singer, singing technique, or vowel.
These classifiers have the same architecture as the combination of FEN, RNNs, and the attention module, and are trained independently from unconverted MELs.

\begin{table*}[ht]
\small
\centering
\begin{tabular}{c|c|c|c|c|c|c|c|c|c|c|c|c|c}
\toprule
\multirow{3}{*}{Strategy}   & \multirow{3}{*}{Model} & \multicolumn{6}{c|}{Effect of Singer Conversion}                                          & \multicolumn{6}{c}{Effect of Technique Conversion}                                       \\ \cline{3-14} 
                            &                        & \multicolumn{2}{c|}{*Singer} & \multicolumn{2}{c|}{Technique} & \multicolumn{2}{c|}{Vowel} & \multicolumn{2}{c|}{Singer} & \multicolumn{2}{c|}{*Technique} & \multicolumn{2}{c}{Vowel} \\ \cline{3-14} 
                            &                        & Before        & After       & Before         & After         & Before       & After       & Before        & After       & Before          & After        & Before       & After       \\ 
                            \midrule \midrule
                            & M0                     & 89.83         & NA             & 90.68          &    NA          & 77.97        &  NA           & 89.83         &   NA          & 90.68           &        NA      & 77.97        &  NA           \\ \midrule
\multirow{3}{*}{C-chunk}                     & M1                     & 80.51         & 63.35       & 83.05          & 75.51         & 77.97        & 69.24       & 80.51         & 75.99       & 83.05           & 54.38        & 77.97        & 72.74       \\
                            & M2                     & 87.29         & 76.95       & 83.90          & 76.99         & 72.03        & 66.78       & 87.29         & 81.92       & 83.90           & 65.82        & 72.03        & 71.33       \\
                            & M3                     & 83.05         & 75.68       & 88.98          & 79.32         & 73.73        & 72.88       & 83.05         & 84.18       & 88.98           & 67.66        & 73.73        & 71.47       \\ \midrule
\multirow{2}{*}{C-sequence} & M2                     & 87.29         & 76.65       & 83.90          & 76.64         & 72.03        & 66.69       & 87.29         & 82.06       & 83.90           & 65.64        & 72.03        & 71.33       \\
                            & M3                     & 83.05         & 75.47       & 88.98          & 79.62         & 73.73        & 72.83       & 83.05         & 84.04       & 88.98           & 67.94        & 73.73        & 72.03       \\
                            \bottomrule
\end{tabular}
\caption{The classification accuracy (\%) derived by the three attribute classifiers, given different models.  * denotes the converted attributes.}
\label{tab:conversion}
\end{table*}

\begin{figure*}[ht]
\centering
\includegraphics[width=0.98\linewidth]{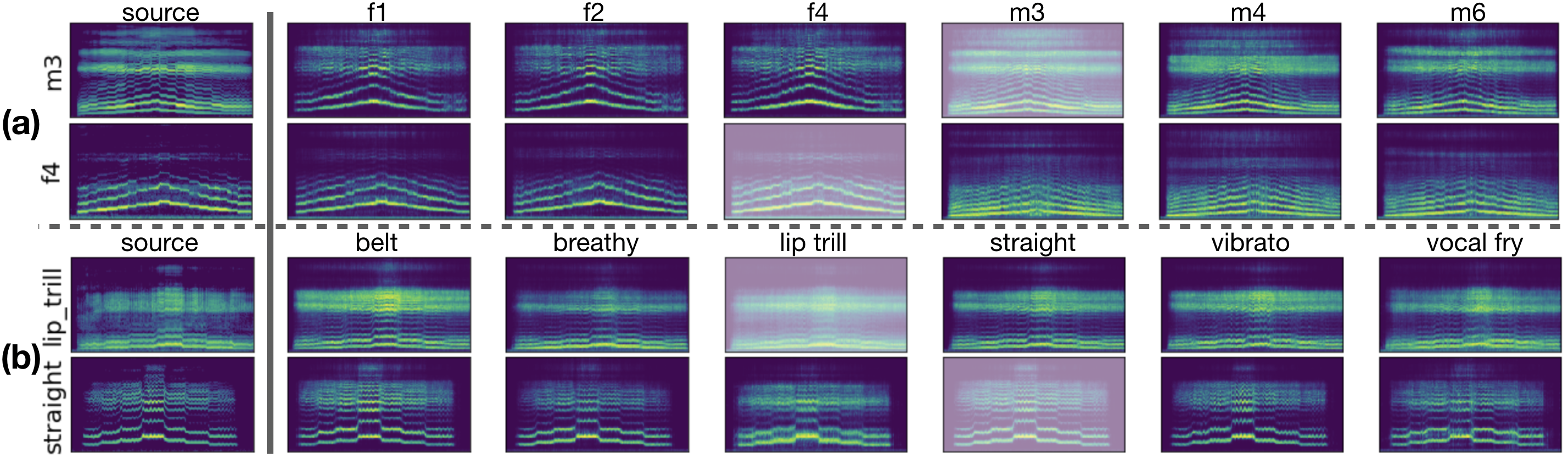}
\caption{Examples of singer conversion (a) and vocal technique conversion (b), converted by the model M3. The first column refers to source, and the rest correspond to different targets. Targets that are the same as the sources are faded.}
\label{fig:spec}
\end{figure*}

\section{Results}
\label{sec:results}
\subsection{Recognizing Attributes from Converted MELs}
We compare three variants of the proposed models:
\textit{M1} denotes the model trained with neither the attention module
nor the the discriminative objectives ($\beta=\gamma=0$).
\textit{M2} is similar to M1 but with $\beta=\gamma=1$.
Finally, \textit{M3} is the model equipped with the attention module.
The performance of M0 servers as the upper bound for classification results using unconverted MELs as input. The results are listed in Table~\ref{tab:conversion}. Higher numbers represent better performance for all cases.

We summarize our findings as follows: 
First, the attributes in the MELs by M2 can be recognized with higher accuracy than those from M1 in all cases;
this supports our belief that including discriminative objectives helps the model to disentangle certain attributes from the input.
Second, results from M2 and M3 are similar,
but M2 is better at singer conversion, 
whereas M3 is slightly better at converting vocal techniques. 
Third, there is no noticeable difference in performance between
the two conversion strategies C-chunk and C-sequence.
Fourth, converting vocal techniques is much more challenging
than converting singer identities as the accuracy drops 
from 90.68\% to below 67.94\% after conversion,
even though the number of techniques is fewer.

\subsection{Visualization}

We visualise some examples of converted MELs in Fig.~\ref{fig:spec}. In the upper panel (a), it is clear that the overall pitch level changes when doing cross-gender conversion.
For technique conversion (b), we can see that converting the lip trill to, e.g., straight, makes the spectrogram less flattened. On the other hand, we can decorate a straight tone by converting it to a bright and energized vocalization as can be seen in straight-belt conversion, or to one with periodic frequency modulation as shown in straight-vibrato conversion. Noticeable effects are also observed when the targets are lip trill and vocal fry.

Despite the change of spectral distribution, the overall pitch contours are retained in all source-target pairs. This hints towards the model's ability to perform many-to-many singer identity and vocal technique conversion.

Conversion at chunk-level enables us to morph from source to target by gradually increasing the conversion vector from 0 to $d\m_{*n}$. As such, we can, e.g., convert a straight tone to gradually express another technique over time.
This has not yet been seen in other existing SVC frameworks, and  
we leave further investigation for future research.

\section{Related Work}
\label{sec:related}

Recent advances in deep learning have brought great success to VC \cite{sun2015voice, Hsu2016vae, Kameoka2018StarGAN};
SVC, on the other hand, have not benefited from it until recently.
SINGAN \cite{sisman2019singan}, an SVC framework based on deep generative models has been proposed to map acoustic features of a source singer to that of a target one;
however, the model is restricted to temporally-aligned singing recordings.
In contrast, \cite{chen2019singing} proposed to combine automatic speech recognition for SVC that is trainable from non-parallel corpora. 
The model, however, only allows for converting a handful of source singers to a single target. 
Recently, an encoder-decoder model is proposed which incorporates a domain confusion network \cite{ganin2016domain} to learn singer-agnostic features \cite{nachmani2019unsupervised}. 

We distinguish ourselves by jointly modeling singer and technique with a principled probabilistic generative model, and conditioning the generation of singing voice on time-dependent latent variables of the aforementioned attributes. To the best of our knowledge, this is the first study on jointly modeling/converting singer identity and vocal technique with a single deep learning model.

\section{Conclusion and Future Work}
\label{sec:final}
We have proposed a flexible framework based on GMVAEs to tackle non-parallel,  many-to-many SVC for singer identity and vocal technique. Audio samples are available from \texttt{https://reurl.cc/oD5vjQ}.
Analyzing the temporal dynamics of the latent variables, as well as accommodating the dependency between singer identity and vocal technique variables will be the focus of our future work.

\vfill\pagebreak



\bibliographystyle{IEEEbib}
\bibliography{strings,refs}

\end{document}